\documentclass[aps,prb,preprint,groupedaddress,showpacs]{revtex4}

\usepackage{graphicx}

\begin{document}


\title{Third order correction to localization in a two-level driven system}


\author{Marco Frasca}
\email[e-mail:]{marcofrasca@mclink.it}
\affiliation{Via Erasmo Gattamelata, 3 \\
             00176 Roma (Italy)}


\date{\today}

\begin{abstract}
A general result is presented on the
lack
of second order corrections 
and on the form of the leading order Floquet quasi-energies in the
high-frequency approximation for a two-level driven system. Then,
by a perturbative approach in the high-frequency approximation
that uses
dual Dyson series 
and renormalization group techniques [M. Frasca, Phys. Rev. B {\bf 68}, 165315 (2003)], we
obtain a third order correction for a sinusoidal driving field
that is
the same obtained by 
Barata and Wreszinski [Phys. Rev. Lett. {\bf 84}, 2112 (2000)], confirming their result that
localization deviates from the zeros of the zero-th order Bessel function when higher order
terms are taken into account. Three different expressions
for this correction are compared. An important consequence of this result is that gives
complete support to the correctness of both methods. Finally, the limitation
of high-frequency approximations is presented by
comparison
with numerical results for a sinusoidal driving field.  
\end{abstract}


\maketitle


In a recent paper by Barata and Wreszinski (BW)\cite{bw} a new method to solve perturbatively the
driven two-level system was presented. These authors showed that a third order correction,
for the case of a sinusoidal driving field, appears to
change
the leading order Floquet quasi energies, in the limit of large frequency,
in such a way to modify
the naive result that one has localization in the two-level system at the zeros of the zero-th
order Bessel function.

This analytical result is in agreement, to a certain extent as we will see, 
with a work by Creffield about the crossing manifolds of the
localization points in this model \cite{cc1,cc2}. A different work by Platero, Brandes and Aguado
\cite{pab} showed that analytical and numerical results agree fairly well taking into
account the result by Barata and Wreszinski.

The question of higher order corrections is a relevant one in view of the needs of workable
two-level systems to be manipulated as qubit in quantum computers. This means that general
results are important to know
the proper way to
drive a two-level system for such aims.

In this paper we present a
several
results on higher order corrections to Floquet quasi-energies
for a driven two-level system. The approach we use has been already presented in \cite{fra1}.
The method given there uses the dual Dyson series
and renormalization group techniques. We improve the computations till third
order showing equivalence with the method of Barata and Wreszinski giving in this way fully
support to both methods in the high-frequency approximation. 

A two-level driven system that is generally considered to implement a qubit by a solid state
device has the form:
\begin{equation}
    H = \frac{\Delta}{2}\sigma_3+g\sigma_1 f(t) \label{eq:H0}
\end{equation}
being $\Delta$ the separation between the two levels, $g$ the coupling constant and 
$f(t)=f(t+T)$ a periodical driving field with period $T$ and $\omega=2\pi/T$
and $\sigma_1$,$\sigma_3$ Pauli matrices. The high-frequency
regime is identified by the condition $\frac{\Delta}{\omega}\ll 1$. Floquet theory can be
applied and from this we know that localization appears at the crossing of the quasi-energies \cite{hg}.
Then,
a proper choice of
the parameters can permit to drive a qubit. The other crucial parameter
for our aims is $z=2g/\omega$ that we will use in the following.

We can prove a general result on the model (\ref{eq:H0}) using the method of
dual Dyson series given in Ref.\cite{fra1}. We introduce the unitary transformation
\begin{equation}
    U_F = \exp\left[-ig\sigma_1\int_0^tdt'f(t')\right]
\end{equation}
that gives the transformed Hamiltonian
\begin{equation}
    H_F=\frac{\Delta}{2}\sigma_3\exp\left[-2gi\sigma_1\int_0^tdt'f(t')\right]. \label{eq:H00}
\end{equation}
By introducing the Fourier series
\begin{equation}
\label{eq:fs}
    \exp\left[-2gi\sigma_1\int_0^tdt'f(t')\right] = \sum_{n=-\infty}^{+\infty}c_n(z)e^{-in\omega t\sigma_1}
\end{equation}
being
\begin{equation}
\label{eq:fc}
    c_n(z)=\frac{1}{T}\int_0^T \exp\left[-in\omega t\sigma_1-2gi\sigma_1\int_0^tdt'f(t')\right]
\end{equation}
where $\sigma_1$ can be treated as a c-number,
it is not difficult to write down the Hamiltonian (\ref{eq:H00}) in the form
\begin{equation}
\label{eq:hh}
    H_F = H_{F0} + \sigma_3\sum_{n\neq 0}a_n(z)e^{-in\omega t\sigma_1}
	+ \sigma_2\sum_{n\neq 0}b_n(z)e^{-in\omega t\sigma_1},
\end{equation}
being $H_{F0}=\sigma_3 a_0(z) + \sigma_2 b_0(z)=\sigma_3 c_0(z)$ the leading order Hamiltonian whose eigenvalues
give the leading order Floquet quasi-energies.
This expression
is obtained by considering that
the coefficients of the Fouries series depend on $\sigma_1$.
Then,
we are able
to derive an estimation of the Floquet quasi-energies through eq.(\ref{eq:fc}) for
any kind of driven two-level model.

We can extend the above result by computing the second-order correction to the quasi-energies
by the method of dual Dyson series \cite{fra1}. The computation with Hamiltonian (\ref{eq:hh})
is quite straightforward giving the following secular contribution to the unitary evolution
\begin{equation}
   U_F^{sec}(t,t_0) =-i[\sigma_3 a_0(z) + \sigma_2 b_0(z)](t-t_0)
   +i\sigma_1\sum_{n\neq 0}\frac{a_n^2(z)+b_n^2(z)}{n\omega}(t-t_0)+\cdots
\end{equation} 
where we can recognize that a second order correction cannot appear when $a_n=\pm a_{-n}$ and
$b_n=\pm b_{-n}$. This is precisely the case for a sinusoidal driving field having
$c_n(z)=J_n(z)$ being $J_n(z)$ the Bessel functions of integer order. This means that eq.(29)
and eq.(30) in Ref.\cite{fra1} 
have not the second order correction and should be corrected by eliminating it. The same
is true for a square wave driving field\cite{cc1,cc2} $f(t)=\Theta(t)-2\Theta(t-T/2)$
in the interval $0\le t < T$. In this case we have the Fourier coefficients
\begin{equation}
    c_n(z)=\frac{z}{\pi}\left[\frac{\sin\pi z}{z^2-n^2}+i\sigma_1\frac{(-1)^n\cos\pi z-1}{z^2-n^2}\right]
\end{equation}
and the leading order Hamiltonian $H_{F0}=\epsilon\left(\sigma_3\cos(\pi z/2)+\sigma_1\sin(\pi z/2)\right)$,
being $\epsilon=\frac{\Delta}{\pi z}\sin(\pi z/2)$ and
$\pm\epsilon$ the leading order Floquet quasi-energies.

A sinusoidal driving field has $f(t)=\cos(\omega t)$. Floquet
theory \cite{hg} gives localization, as a first approximation,
for $J_0(z)=0$.
One may ask if this result is true at any order or just as a first order approximation.
Barata and Wreszinski obtained that 
a third order correction given by
\begin{equation}
    BW(z) = -\frac{\Delta^3}{4\omega^2}\sum_{n_1,n_2}
	\frac{J_{2n_1+1}(z)J_{2n_2+1}(z)J_{-2(n_1+n_2+1)}(z)}{(2n_1+1)(2n_2+1)}
\end{equation}
that is not generally zero at the zeros of $J_0(z)$. In a successive work Barata and Cortez \cite{bc}
proved that the same correction could be written as
\begin{equation}
    BC(z) = -\frac{\Delta^3}{4\omega^2}\sum_{n_1\neq 0,n_2\neq 0}\frac{J_{n_1}(z)J_{n_1-n_2}(z)J_{n_2}(z)}{n_1n_2}.
\end{equation}

Our method given in Ref.\cite{fra1} needs the computation of the third order term 
of the unitary evolution by dual Dyson series
\begin{equation}
    T_3 = i\frac{\Delta^3}{8}\sigma_z\int_{t_0}^tdt_1\int_{t_0}^{t_1}dt_2\int_{t_0}^{t_2}dt_3
	\sum_m\sum_n\sum_p J_m(z)J_n(z)J_p(z) e^{-i\omega\sigma_x(mt_1-nt_2+pt_3)}
\end{equation} 
starting from a generic initial time $t_0$ and being $J_n$ Bessel functions of integer order.
Without repeating all the procedure here, our approach gives the third order correction
\begin{equation}
    F(z) =-\frac{\Delta^3 J_0(z)}{4\omega^2}\sum_{n\neq 0}\frac{J_n^2(z)}{n^2} 
	-\frac{\Delta^3}{8\omega^2}\sum_{n_1\neq 0,n_2\neq 0,n_1\neq -n_2}
	\frac{J_{n_1}(z)J_{n_1+n_2}(z)J_{n_2}(z)}{n_1n_2}.
\end{equation}
that yields the complete term with respect to eq.(31) in Ref.\cite{fra1}. We have given
explicitly the first term with $J_0(z)$ that recovers the small $z$ limit. This limit is
recovered by all the expressions $BW(z)$, $BC(z)$ and $F(z)$ as it should be. 
As already pointed out by Barata and Cortez \cite{bc}, it is not possible to prove directly
equivalence of this formulas by analytical means. The relationship does not depend on the
properties of the Bessel functions. So, we need to content ourselves of a direct numerical check here.

\begin{figure}[t,b,p]
\begin{center}
\includegraphics[height=0.5\textwidth,width=1\textwidth]{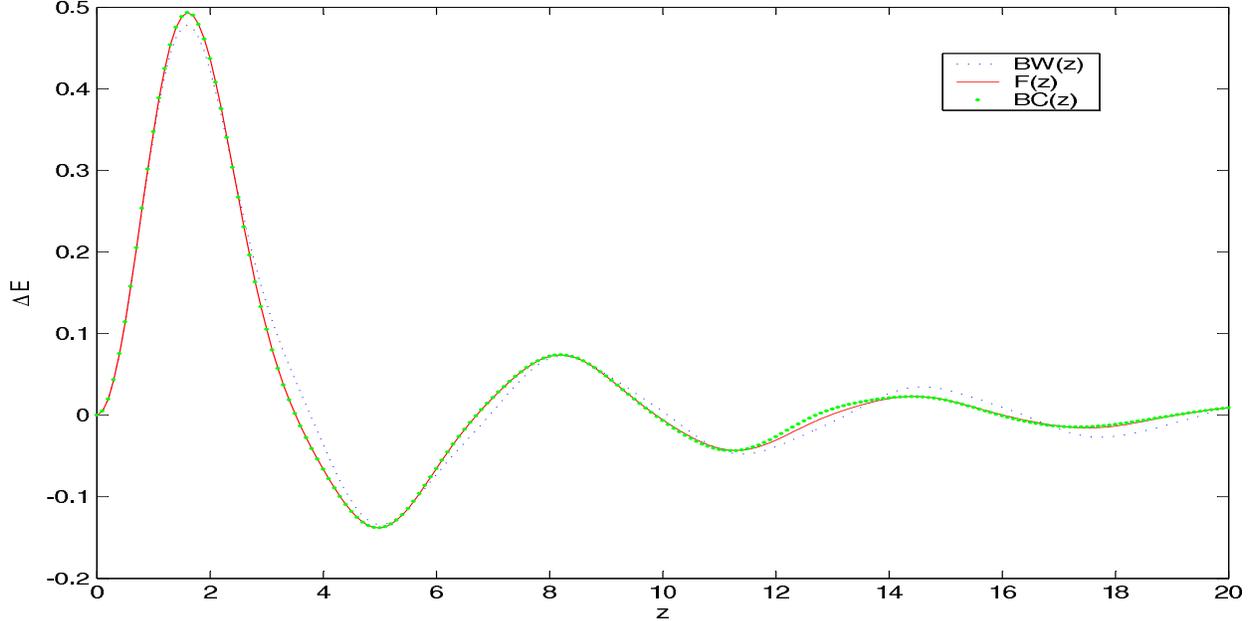}
\caption{\label{fig:fig1} (Color online) The three terms, normalized by the factor $\frac{\Delta^3 J_0(z)}{4\omega^2}$, 
are given here in red (solid) for $F(z)$,
green (points) for $BC(z)$ and blue (dotted) for $BW(z)$. The red and green curves are almost identical.}
\end{center}
\end{figure}

In fig.\ref{fig:fig1} we have compared the three expressions above, normalized by the factor
$\frac{\Delta^3 J_0(z)}{4\omega^2}$ and the result is absolutely satisfactory.

So, the BW method and our method give both the same result supporting each other with
respect to correctness as should be expected. To use one or the other becomes just a matter of
taste. The fact that two different analytical approaches, in the limit of large frequency,
give the same result permits us to conclude that, also by this way, it is definitely
proved that zeros of $J_0(z)$ are just approximately localization points .

\newpage

\begin{figure}[t,b,p]
\begin{center}
\includegraphics[height=0.5\textwidth,width=1\textwidth]{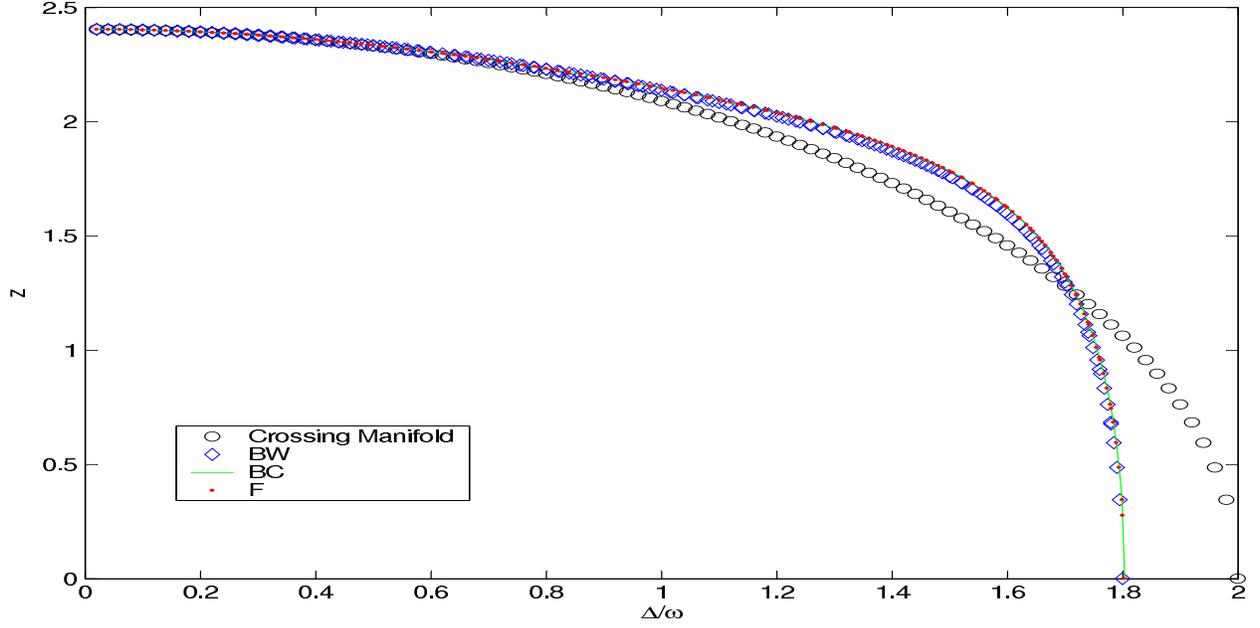}
\caption{\label{fig:fig2} (Color online)
 The three terms, $F$ in red (dotted), $BC$ in green (solid) and $BW$ in blue (diamonds)
 are compared with the numerical results on crossing manifolds in black (circles) given in
 Ref.\cite{cc1}.}
\end{center}
\end{figure}

One may ask how far this third order corrections can be pushed in the ratio $\Delta/\omega$.
More generally, it is rather interesting to understand the limitations of such computations.
The answer can be obtained from fig.(\ref{fig:fig2})
where is seen that for values of the ratio $\Delta/\omega\approx 0.5$ all the high-frequency
approximations begin to fail demanding for higher order terms.

In order to give a proper picture of the problems encountered in Ref.\cite{fra1} we note
as the series we obtained in it for the Floquet quasi-energies till third order was not
correct [eqs.(29),(30) and (31) in Ref.\cite{fra1}]. 
Rather we have shown here that no second order correction appears and that the
third order correction deviates from the naive leading order result of localization at
the zeros of the zero-th order Bessel function. These corrections make our approach
completely equivalent, for the results, to the method conceived by Barata and Wreszinski
\cite{bw}. It should be said that our method makes the computation of Floquet eigenstates
and quasi-energies a completely algorithmic matter, relying all the calculations on a very
simple recipe.

In conclusion we have given general results for a driven two-level model giving a
method to compute high-frequency leading order Floquet quasi-energies. Besides, we have
proved fully equivalence between Barata and Wreszinski and our method to compute a high-frequency
perturbation series. Finally, the limitation of these methods has been pointed out by comparing them
with a numerical computation.

\newpage

\begin{acknowledgments}
It is a pleasure to acknowledge the precious help by Charles Creffield to improve the
content of this paper. Fig.(\ref{fig:fig2}) has been possible by the data 
kindly given by him.
\end{acknowledgments}


\end{document}